# Mössbauer spectroscopy of
# short-range ordered Fe-Cr alloys

G. Apostolopoulos, K. Mergia, and S. Messoloras

*Institute of Nuclear and Radiological Science and Technology, Energy, and Safety*

*National Centre for Scientific Research "Demokritos"*

*15310 Aghia Paraskevi Attikis, Greece*

(Dated: 27 April 2012)

## Abstract

Short-range ordering (SRO) in Fe-Cr has been the subject of a number of recent experimental and theoretical investigations, as ordering effects are significant for the phase stability of this technologically important alloy. Recently, discrepancies regarding the SRO behavior of Fe-Cr have been reported between results obtained with Mössbauer spectroscopy and diffuse neutron scattering, respectively. As the methodology for reducing SRO parameters from Mössbauer spectra is indirect and relies on the determination of a large number of parameters, here a new method for directly connecting the Mössbauer effect with SRO is presented. The method is verified using synthetic spectra derived from Monte-Carlo simulated alloy structures with different SRO parameters and subsequently it is applied to experimental data obtained from Fe-Cr alloys. Agreement with diffuse neutron scattering is found at low Cr concentrations; however, some discrepancies still remain for more concentrated alloys and possible reasons for these are discussed.

## I.    INTRODUCTION

The Fe-Cr binary alloy system forms the base of a very important class of engineering materials, ferritic stainless steels, which combine good mechanical properties, corrosion resistance, and also resistance to nuclear radiation environments. Fe-Cr alloys exhibit a complex behavior due to the interplay between configurational and magnetic energies. The low temperature/low Cr concentration region of the phase diagram, where the alloy is in the ferromagnetic bcc state, is still under investigation. Mirebeau et al. [1, 2] observed by means of diffuse neutron scattering (DNS) that at low Cr concentrations the alloy exhibits a strong tendency for short-range ordering (SRO), as indicated by a negative value of the Warren-Cowley SRO parameter. At higher concentrations the behavior changes to Cr clustering and the Warren-Cowley parameter becomes positive as the alloy approaches the solubility limit. The SRO inversion has been recently correlated [3, 4] with *ab-initio* electronic structure calculations [5, 6] showing a change of sign in the Cr mixing enthalpy at the same concentration region where the SRO inversion occurs. These findings are of particular importance for the phase stability of Fe-Cr and for the evolution of the microstructure and may have serious implications for the hardening and embrittlement behavior of ferritic steels under irradiation [7].

Mössbauer spectroscopy is a powerful technique that can provide information on the local chemical environment around Fe atoms and has been frequently employed for the study of ferromagnetic iron alloys [8, 9, 10, 11, 12]. In a recent publication Dubiel & Cieslak [13] attempted to obtain the Fe-Cr SRO parameters from their Mössbauer spectra. However, their results are not in agreement with DNS experiments [1, 2]. According to their interpretation, the SRO parameter, averaged over the first two nearest neighbor shells, does not change sign as a function of Cr concentration. Instead, the averaged SRO parameter can attain negative or positive values depending on the metallurgical condition of the sample.

As already noted by Schwartz and Asano [14], the extraction of SRO parameters from Mössbauer spectra presents fundamental difficulties related to the type of information obtained from the experiment. Mössbauer spectroscopy yields the probability $P_{m_1 m_2}$ that an iron atom has $m_1$ solute atoms in nearest neighbor sites and $m_2$ solute atoms in next-nearest neighbor sites. In order to obtain the SRO parameters from these probabilities, a full set of $P_{m_1 m_2}$ is required; however, experimentally only a limited number of $P_{m_1 m_2}$ values can be accurately measured.

In view of these facts, this paper presents a novel procedure for unraveling Mössbauer spectra of ferromagnetic iron alloys, which yields directly the Warren-Cowley SRO parameters and the alloy concentration. It is based on Probability Variation Method (PVM) introduced by Clapp [15, 16], which allows one to obtain configuration probabilities of an atomic cluster given the SRO parameters. Apart from directly providing SRO information, the new method has the following additional advantages: (i) it can be efficiently implemented numerically and incorporated in least-square fitting schemes; (ii) it reduces significantly the number of adjustable parameters, making the analysis of Mössbauer spectra much more robust. The new method is initially verified by its application to simulated Mössbauer spectra, obtained from Monte-Carlo generated alloy configurations exhibiting different SRO parameters; subsequently, it is employed in the analysis of experimental spectra from Fe-Cr alloys of different concentration and metallurgical condition.

## II.   THE MÖSSBAUER EFFECT IN IRON BASED FERROMAGNETIC ALLOYS

The Mössbauer spectra of pure ferromagnetic iron consist of the characteristic resonance sextet, which reflects the splitting of the ground and first excited states of $^{57}$Fe in the presence of the hyperfine magnetic field (HMF), $H_{\mathrm{Fe}}$, which has a magnitude of $330 \mathrm{~kG}$. The most important contribution to the HMF is due to the Fermi contact interaction [17] between the nuclear magnetic moment and the electronic spin density at the position of the nucleus. This effect involves only the core and valence $s$ electrons, which have non-vanishing density at the origin. In iron the spin unbalance of the magnetic-moment carrying $3d$ orbitals is mediated to the $s$ states through exchange and polarization effects. This results in the un-pairing of $s$ states at the origin and in the appearance of the HMF. In iron alloys the presence of solute atoms perturbs the HMF on neighboring iron atoms, either by affecting their magnetic moment or by causing a redistribution of their $4s$ states [18]. Thus, the HMF depends on the configuration of solutes in the immediate neighborhood of an iron atom. This is reflected in the Mössbauer spectra which exhibit much more structure than those of pure iron, especially at the two outer peaks of the six-line pattern. For an alloy containing $N$ Mössbauer nuclei and if $\Delta H_i$ denotes the HMF change for the $i$-th nucleus, then the resonant absorption measured as a function of source velocity $v$ is

$$a(v) = \frac{1}{N} \sum_i^N \sum_{j=1}^6 h_j L\left[v - \delta_i - z_j\left(H_{\mathrm{Fe}} + \Delta H_i\right); \gamma\right]$$
$$= \sum_k P_k \sum_{j=1}^6 h_j L\left[v - \delta_k - z_j\left(H_{\mathrm{Fe}} + \Delta H_k\right); \gamma\right] \qquad (1)$$

where $L(x; \gamma)$ denotes the Lorentzian nuclear resonance profile of width $\gamma$, $h_i$ and $z_i$ define the amplitude and position of individual resonance lines within a sextet, respectively, and $\delta$ is the isomer shift. The values of $\gamma$, $h_i$ and $z_i$ are constants characteristic of the $^{57}$Fe Mössbauer resonance while the hyperfine parameters $\Delta H_i$ and $\delta_i$ are determined by the specific solute configuration around the Mössbauer atom. The first summation in equation (1) is over all $^{57}$Fe nuclei whereas the second is over all different configurations and $P_k$ is the probability of a specific configuration. As interpretation of the experimental Mössbauer results based on equation (1) is unattainable, simplifications are needed. An obvious simplifications of (1) is the assumption that the additivity of the HMF shifts established in ordered alloys (e.g. Fe$_3$Si and Fe$_3$Al, see [19]) applies to all systems. There is also a strong indication that for random FeCr alloys produced by fast quenching [12] or for alloys with high Cr content [20] the additivity still holds to a good approximation. The additivity of the HMF shifts is expressed as

$$\Delta H_k = \Delta H_{m_1 \ldots m_l} = \sum_{n=1}^l m_n \, \Delta H_n \qquad (2)$$

where $m_n$ is the number of $n$-th order nearest neighbor sites occupied by solute atoms and $\Delta H_n$ is the contribution to the shift of a single solute atom at a $n$-th nearest neighbor position. The probability $P_k$ in the case the configuration is expressed in occupation numbers of the neighboring cells is denoted by $P_{m_1 \ldots m_l}$.

In a dilute Fe alloy a random distribution of impurities may be assumed and therefore the different probabilities $P_{m_1 \ldots m_k}$ can be expressed analytically, by means of the binomial distribution, or numerically [21] and the hyperfine fields are then derived by a least square fit of equation (1) to the experimental data. In most of the work performed on dilute alloys [8, 9, 10] the analysis has been restricted to the first two nearest neighbor shells [$l \leq 2$ in eq. (2)], the HMF shifts caused by higher order neighbors being considered negligible. However, other workers have included higher order coordination shells [22, 21] in their interpretation. In recent *ab-initio* calculations of the HMF shifts caused by $3d$ impurities, in the single impurity limit [23] and for a fixed concentration of 3.7% [24], it has been shown that, at least for Cr, the contributions from nearest neighbor shells higher than the second are much smaller and their effect cannot be observed within the resolution of Mössbauer spectroscopy. Furthermore, the values of $\Delta H_1$ and $\Delta H_2$ reported in [23, 24] for $3d$ solutes are in good agreement with experimental results from dilute alloys.

Turning to concentrated alloys, the analysis of Mössbauer spectra based on equations (1) and (2) suffers by some important limitations. First, the assumption of random solute distribution, whenever it applies is questionable. Even if we assume that the high temperature thermodynamic state of the sample is frozen by fast quenching a degree of short range order will be present in many cases even at high temperatures. Further, the assumption of the additivity of HMFs [eq. (2)], although there is currently no strong experimental evidence against it, requires a theoretical justification, especially in high concentration alloys in which the average number of solute atoms around Fe is large. Despite this lack of justification, the assumption of additivity offers a significant advantage: it provides a means for calculating the HMF in a large number of different configurations by just a few parameters. A similar tool is missing for the configuration probabilities, which are typically left as a large number of freely adjustable parameters, constrained only by normalization, in a least-squares analysis of experimental spectra. The methodology developed in this work actually provides such a tool; it diminishes the number of the required parameters to only those needed to describe the configurational state of the alloy, i.e., the SRO parameters. In addition, the alloy concentration can be considered as an additional parameter to be fitted. As the Mössbauer spectrum is received by a small sample volume prepared by different means from a bulk sample its composition might differ from that of the parent material and thus it is beneficial the solute concentration to be determined. This is essential after a heat treatment or irradiation of an alloy when a percentage of solute atoms have agglomerated.

## III.   ASSOCIATION OF MÖSSBAUER SPECTRA WITH SHORT RANGE ORDER

In equation (1) we have connected quite generally the Mössbauer spectrum with the probability $P_k$ of the atomic configuration around a Mössbauer atom. Employing the terminology of cluster expansions [25], the configuration $k$ of a cluster of $M$ atoms is uniquely described by a vector of occupation numbers $\mathbf{c}_k = \left\{ c_1^\mu c_2^{\mu'} \dots c_M^{\mu''} \right\}$, where $c_i^\mu$ takes on the value of 1 if site $i$ is occupied by an atom of species $\mu$, or zero, otherwise. The probability $P_k$ entering eq. (1) for the occurrence of a particular atomic configuration is then given by a multisite correlation function

$$P_k = P(\mathbf{c}_k) = \left\langle c_1^\mu c_2^{\mu'} \dots c_M^{\mu''} \right\rangle \tag{3}$$

regarding the occupation of a number of sites $M$ constituting the environment of the Mössbauer atom. In the last equation $\left\langle \dots \right\rangle$ denotes an ensemble average. Site occupation correlations play a major role in the formulation of configurational alloy thermodynamics (for a review see [25, 26]). In this framework the alloy Hamiltonian is expanded in terms of atomic clusters, i.e., geometrical combinations of lattice sites. The expansion coefficients are called effective cluster interactions and are obtained by first-principles calculations performed on a limited set of configurations. The success of cluster expansion methods owes to their rapid convergence properties. Typically only a few terms are needed to capture the physical properties of a system. Of particular importance are expansions truncated to the lowest order clusters, i.e., to atomic pairs. Pair correlations are usually the strongest ones and moreover they are experimentally accessible and can be used directly for the validation of theoretical models. The Warren-Cowley SRO parameters $\alpha_n$ deduced from diffuse scattering experiments [27] are directly related to pair correlations. For a cubic binary alloy with constituents A and B

$$\alpha_n = 1 - \frac{\left\langle c_0^A c_n^B \right\rangle}{c_A c_B}, \tag{4}$$

where $\left\langle c_0^A c_n^B \right\rangle$ is the correlation between an atom at the origin and its $n$-th order nearest neighbor. Furthermore $c_\mu = \left\langle c_i^\mu \right\rangle$ is the concentration of species $\mu$ with $\mu = $ A or B and $c_A + c_B = 1$.

Recently it has been proved rigorously that when the cluster expansion Hamiltonian contains only pair interactions, all higher order correlations are uniquely determined by the pair correlations [28]. In the particular case of Fe-Cr alloys, recent work on the cluster expansion, based on detailed first-principles electronic structure calculations, shows that effective pair interactions are the most significant ones with smaller contributions from higher order clusters [3, 29, 30].

Based on the above, we propose a model for the Mössbauer effect in short-range ordered Fe-Cr alloys based on the following assumptions:

i) The $^{57}$Fe hyperfine parameters are determined by the distribution of Cr solute atoms only within the first two nearest neighbor coordination shells. Effects of higher order shells are neglected. The relevant configuration probabilities $P_k$ entering eq. (1) are those of the atomic cluster composed of a central atom and the first two nearest neighbor shells. In the bcc lattice this "Mössbauer cluster" contains $M = 15$ atoms.

ii) The configurational energy of the alloy is dominated by pair interactions. Thus, the $P_k$ of the Mössbauer cluster are determined by pair correlations alone. This is expressed as

$$P_k = P_k\left(c, \boldsymbol{\alpha}\right) \tag{5}$$

where $c$ is the alloy concentration and $\boldsymbol{\alpha}$ denotes a set of SRO parameters representing the pair correlations as defined in eq. (4). The Mössbauer cluster contains pairs with interatomic distances up to the 6$^{th}$ coordination shell, thus $\boldsymbol{\alpha}$ is composed of the first 6 SRO parameters.

iii) Changes in hyperfine parameters induced by Cr solutes are additive. Consequently, only the number of Cr atoms as first and second nearest neighbors matter and not the exact sites they occupy. Thus, the quantity of interest for Mössbauer spectroscopy is the conditional probability $P_{m_1 m_2}$ for the occurrence of $m_1$ first and $m_2$ second order Cr nearest neighbors, given an Fe atom is at the center of the cluster. If $\mathcal{K}(m_1, m_2)$ denotes the subset of configurations that fulfill these conditions then

$$P_{m_1 m_2}\left(c, \boldsymbol{\alpha}\right) = \frac{\displaystyle\sum_{k \in \mathcal{K}(m_1, m_2)} P_k\left(c, \boldsymbol{\alpha}\right)}{1 - c}, \tag{6}$$

where the denominator accounts for the probability of the central site being occupied by an Fe atom.

In the above chain of arguments the only missing link is the actual form of the functions $P_k\left(c, \boldsymbol{\alpha}\right)$. Only in the limit of a random alloy, where all SRO parameters tend to zero, can an analytical expression be found. In this case the $c_i^\mu$ are uncorrelated and thus

$$\lim_{\boldsymbol{\alpha} \to 0} P_k(c, \boldsymbol{\alpha}) = \lim_{\boldsymbol{\alpha} \to 0} \left\langle c_1^\mu c_2^{\mu'} \ldots c_M^{\mu''} \right\rangle =$$
$$\left\langle c_1^\mu \right\rangle \left\langle c_2^{\mu'} \right\rangle \cdots \left\langle c_M^{\mu''} \right\rangle = c^{m_1 + m_2} \left(1 - c\right)^{M - m_1 - m_2}.$$

Furthermore, the number of configurations belonging to the subset $\mathcal{K}(m_1, m_2)$ is equal to the number of ways the $(m_1, m_2)$ Cr atoms can be distributed in the available positions in the respective coordination shells. Finally, $P_{m_1 m_2}$ can be written in terms of the binomial distribution

$$\lim_{\boldsymbol{\alpha} \to 0} P_{m_1 m_2}\left(c, \boldsymbol{\alpha}\right) = B\left(m_1; M_1, c\right) B\left(m_2; M_2, c\right),$$
$$B\left(m_i; M_i, c\right) = \frac{M_i!}{m_i!\left(M_i - m_i\right)!} c^{m_i}\left(1 - c\right)^{M_i - m_i}, \quad M_1 = 8, M_2 = 6 \tag{7}$$

The lack of a similar analytical expression for $P_{m_1 m_2}$ in the presence of short-range order has hindered so far the wide-spread use of the Mössbauer technique for the study of ordering effects in alloys. It should be mentioned here the approach employed in previous work [13, 31, 32] for connecting the SRO parameters with the Mössbauer spectra. First the $P_{m_1 m_2}$ are determined by fitting of $\left(M_1 + 1\right)\left(M_2 + 1\right) = 63$ parameters (or a fraction fewer by ignoring the less probable ones) and the apparent concentration $\overline{c}$ of solute atoms in the neighboring shells is computed as $\overline{c}_{1(2)} = M_{1(2)}^{-1} \sum_{m_1, m_2} m_{1(2)} P_{m_1 m_2}$. The SRO parameters are then obtained by the well-known relation $\alpha_{1(2)} = 1 - \overline{c}_{1(2)} / c$. This approach being criticized for numerical instabilities [14] has two shortcomings from its onset as it assumes that all $P_{m_1 m_2}$ variables are strongly uncorrelated and that the Mössbauer spectrum contains all the necessary information for the determination of a large number of parameters. We claim that the $P_{m_1 m_2}$ variables are weakly dependent on higher order correlation functions and further that the Mössbauer spectra do not have the required resolution for the unequivocal determination of so many parameters. This assertion will be substantiated below by fitting experimental data employing essentially only one SRO parameter.

It has already been noted earlier [14] that in lieu of analytical formulas for $P_{m_1 m_2}$ one should turn to numerical methods. As a first step in this direction the Monte-Carlo (MC) technique is employed to simulate Mössbauer spectra of short-range-ordered Fe-Cr. A model alloy configuration is constructed numerically so that it complies with a set of specified SRO parameters. All higher order correlations present in this model alloy are consistent with a system Hamiltonian containing only pair interactions. The adopted Monte-Carlo procedure of constructing a binary alloy structure with defined concentration and SRO parameters is that described in [33]. In short, starting from a purely disordered configuration, pairs of A-B atoms are selected at random and their positions are exchanged. If after the exchange the pair correlations in the lattice are closer to the desired values, the exchange is retained; otherwise it is rejected. This is repeated until the required pair correlations are established. Simulations were performed on model bcc lattices of $2 \times 10^6$ atoms. These MC calculations will serve as the base for testing the developed numerical procedure for deriving the SRO parameters from experimental Mössbauer spectra and which is going to be discussed in the next section.

Mössbauer spectra of short-range ordered Fe-Cr alloys have been simulated in the following way: for a randomly chosen Fe atom of the simulated lattice the first and second neighbor Cr atoms are counted and the corresponding Mössbauer spectrum is calculated using equation (1) and (2). The relative HMF shifts are set equal to values typical of Fe-Cr alloys, $\Delta H_1 / H_{Fe} = -10\%$, $\Delta H_2 / H_{Fe} = -7\%$. The shifts are negative due to the fact that Cr atoms couple anti-ferromagnetically to the iron lattice and thus tend to reduce the HMF of neighboring iron nuclei. Other parameters as, e.g., Mössbauer linewidth and isomer shifts also take on typical values. The procedure is repeated for a large number of Fe atoms, the spectra are summed and normalized. A computer-generated random Gaussian noise signal is added to account for the typical counting errors in a real experiment.

In fig. 1 the MC simulated Mössbauer spectra for a Fe-15 at. % Cr alloy and for three different SRO parameters are shown. The six-line pattern of pure Fe [spectrum (a)] is also depicted for comparison. As seen in the figure, the peaks of the six-line pattern split into a number of satellites, which correspond to the different Cr configurations around an Fe atom. The splitting is more pronounced in the $1^{st}$ and $6^{th}$ peak. The outermost satellites of these peaks are due to isolated iron atoms with zero Cr neighbors; these have almost the same HMF as in pure Fe. As the number of Cr neighbors increases, the HMF decreases and gives place to the inner satellites. Spectrum (b) corresponds to a disordered alloy configuration and the relative intensities of the satellite peaks are in accordance with the binomial distribution. In spectrum (c) a SRO configuration is assumed with $\alpha_1 = \alpha_2 = \alpha = -0.05$ and $\alpha_n = 0$ for $n \geq 3$ while in (d) the first two parameters are positive $\alpha_1 = \alpha_2 = \alpha = 0.05$. The relative intensities of satellites depend on the value of the SRO parameter. For example, the intensity of the outermost satellite in spectrum (c) is reduced with respect to the disordered alloy (b) while the intensity of the inner satellites is increased. On the other hand, in spectrum (d) the intensity of the outermost satellites is increased at the expense of the inner ones. This behavior can be understood qualitatively as follows: in case (c), where $\alpha < 0$, Cr atoms tend to distance themselves from one another, favoring the presence of iron atoms between them; thus, the percentage of iron atoms having one or two solute neighbors is increased; for $\alpha > 0$, spectrum (d), Cr atoms tend to form clusters and are effectively removed from the alloy matrix; thus the apparent alloy concentration is reduced and the percentage of iron atoms having one or more solute neighbors is decreased. The sensitivity of the relative satellite intensities on the value of the SRO parameter shows that Mössbauer spectroscopy is well suited for the study of ordering effects.

With the same MC procedure the probabilities $P_{m_1 m_2}(c, \boldsymbol{\alpha})$ for the occurrence of the $(m_1, m_2)$ first and second neighbor configuration can be derived by mere counting of the frequency of their occurrence within the model crystals. In fig. 2 the first four $P_{m_1 m_2}$ with $m_1, m_2 \leq 1$ are shown as a function of solute concentration for $0 \leq c_B \leq 0.2$. Upper triangles depict the MC calculated $P_{m_1 m_2}(c, \boldsymbol{\alpha})$ with $\alpha_1 = \alpha_2 = \alpha = -0.05$ while lower triangles correspond to $\alpha_1 = \alpha_2 = \alpha = 0.05$. The $P_{m_1 m_2}$ of the random alloy given by the binomial distributions of eq. (7) are also shown as a continuous line. The behavior of $P_{00}$ depicted in fig. 2(a) is monotonous with respect to $\alpha$. When $\alpha > 0$ the probability $P_{00}$ is always increased with respect to the random alloy while for $\alpha < 0$ it is always decreased, regardless of alloy concentration. Higher-order probabilities shown in fig. 2(b)-(d) exhibit a more complex behavior as a function of concentration. For example, $P_{10}$ with $\alpha > 0$ is below the random alloy value for $c < 0.12$ but goes above the random alloy curve for higher concentration. This complex behavior is due to the intricate interplay between the probability for the occurrence of, e.g., A-B pairs, which increases with concentration, and the correlations between these pairs which are governed by the value of $\alpha$.

## IV.   DETERMINATION OF SRO PARAMETERS FROM THE MÖSSBAUER SPECTRA

The MC procedure described above is an excellent approach for calculations or qualitative discussion, however, it is very inefficient for the reduction of the different parameters from experimental data as the best approach is the least squares fit. In this section we introduce a novel numerical procedure for obtaining the configuration probabilities $P_{m_1 m_2}(c, \boldsymbol{\alpha})$ that is efficient enough to be included in a non-linear fitting scheme. This allows for the determination of the alloy concentration and of the SRO parameters through the analysis of experimental Mössbauer spectra.

Before proceeding to present this new procedure we note here briefly the exact functional form that is used for fitting to experimental spectra. In order to take into account thermal vibrations, local strains and experimental resolution effects the discrete summation of (1) is replaced by the convolution of the Lorentzian with a field distribution function having Gaussian form [34]. A single Mössbauer line is therefore written as

$$\int dH \, G(H - H_{m_1 m_2}; \sigma) L(v - \delta - z_i H; \gamma) = V(v - \delta - z_i H_{m_1 m_2}; \gamma, |z_i| \sigma), \qquad (8)$$

where $G(x; \sigma)$ is a Gaussian function centered at

$$H_{m_1 m_2} = H_{Fe} + \Delta H_0 + m_1 \Delta H_1 + m_2 \Delta H_2 \qquad (9)$$

and of width $\sigma$ ; $V(x; \gamma, \sigma)$ denotes the Voigt function. Finally, the expression to be fitted to experimental data by standard non-linear least squares fit is given by

$$a(v) = \sum_{m_1=0}^{8} \sum_{m_2=0}^{6} \sum_{i=1}^{6} h_i \, P_{m_1 m_2}(c, \boldsymbol{\alpha}) \, V(v - \delta - z_i H_{m_1 m_2}; \gamma, |z_i| \sigma). \qquad (10)$$

The numerical evaluation of $P_{m_1 m_2}(c, \boldsymbol{\alpha})$ is based on the Probability Variation Method (PVM) originally proposed by Clapp [15, 16]. We start by first enumerating all possible configurations $k$ of the 15-atom Mössbauer cluster. Lattice symmetry reduces the number of distinct configurations from $2^{15}$ to $K = 852$. If $W_k$ denotes the number of crystallographically equivalent configurations and $P_k$ is the corresponding occurrence probability, then the normalization condition reads as

$$\sum_{k=1}^{K} W_k P_k = 1. \qquad (11)$$

In order for the $P_k$ to reflect the alloy concentration, the following sum rule has to be obeyed

$$\left\langle c_i^B \right\rangle = \sum_{k=1}^{K} \left[ c_i^B \right]_k W_k P_k = c_B, \qquad i = 1, \ldots, M \qquad (12)$$

Here $\left[ c_i^B \right]_k$ denotes the value of $c_i^B$ within the $k$-th cluster configuration. A set of similar sum rules may be written for the pair correlations

$$\left\langle c_i^A c_j^B \right\rangle = \sum_{k=1}^{K} \left[ c_i^A c_j^B \right]_k W_k P_k = c_A c_B \left( 1 - \alpha_n \right) \qquad (13)$$

where the indexes $(i, j)$ run over all possible $n$-th order pairs within the cluster. The right hand side of the last equation follows directly from eq. (4) and provides the link between the cluster configuration probabilities and the Warren-Cowley SRO parameters. We note that the atoms in the Mössbauer cluster form a total of $M(M-1)/2 = 105$ pairs with coordination up to the $6^{\text{th}}$ shell.

The equations (11)-(13) do not form a complete set, i.e., they are not sufficient to provide a full description of $P_k$. The idea of Clapp was that an approximate set of $P_k$ can be obtained by maximizing the entropy measure

$$S = -\sum_{k=1}^{K} W_k P_k \log P_k \tag{14}$$

subject to the constraints imposed by the sum rules (11)-(13). The physical basis behind this proposition is the following: if pair interactions and, consequently, pair correlations are dominant, then eq. (13) ensures that the entropy maximization is done while keeping the configurational energy of the system constant. This is equivalent to maximizing the entropy of a microcanonical ensemble. It is noted that (14) is only an approximation to the true entropy of the alloy, however, Clapp [16] has argued that the maximum of (14) will lie very close to the maximum of the exact entropy.

The information required to implement the PVM includes, for each distinct configuration $k$, the multiplicity $W_k$, the number of B atoms, and the number of $n$-th order A-B pairs with $n \leq 6$. These are easily obtained by inspection. The constraint entropy maximization is accomplished by the standard method of Lagrange multipliers as already described by Clapp [16]. The maximization problem is thus converted to the solution of a system of non-linear equations. Solution by standard numerical methods yields the cluster configuration probabilities $P_k(c, \boldsymbol{\alpha})$ and with the aid of eq. (6) the conditional probability $P_{m_1 m_2}(c, \boldsymbol{\alpha})$. The numerical procedure has been implemented in a special software routine [35] which is used in combination with eq. (10) for fitting to experimental data.

The PVM might be criticized on the ground that it utilizes entropy maximization and not minimization of the total free energy of the system. It could be argued that by specifying the SRO parameter only clusters contributing to the energy minimization are selected and not all possible cluster configurations. The use of only an approximate entropy measure may be another concern. To test these arguments we compare results derived from PVM with the MC simulations discussed in the previous section. We first analyze the three simulated Mössbauer spectra of fig. 1 using the methodology discussed above and the corresponding software. The fitted spectra are shown as continuous lines in fig.

1. Also shown is the difference between MC simulation and the fitted spectrum. We find excellent agreement between fitted lines and simulated spectra. The values of the different parameters input to the MC calculations and the corresponding results from the leat-squares minimization are listed in Table I. As observed in the table, all fitted parameters are very close to the input values. Specifically, the maximum absolute deviations are: 0.17 at. % for the alloy concentration, 0.009 for the SRO parameter, and 0.06 % for the relative HMF shifts.

Next we compare the $P_{m_1 m_2}(c, \boldsymbol{\alpha})$ obtained by MC in section III as a function of concentration (shown with up and down triangles is fig. 2) with the ones derived by the PVM. The latter are depicted in fig.2 by dashed lines. Again very good agreement is seen between MC and PVM results both qualitatively and quantitatively. In some cases small deviations of the order of a few percent can be observed between the values calculated by the two methods. This is attributed to the approximate form used by the PVM for the entropy measure. In general these results show that the PVM can be used with confidence for the calculation of configuration probabilities relevant to the Mössbauer effect. Further extensive simulations over a wide range of concentrations and Warren-Cowley SRO parameters have shown the following:

i)    Mössbauer spectra are not sensitive to SRO parameters $\alpha_n$ with $n > 2$. Although the Mössbauer cluster contains pairs of higher coordination, the effect of the relevant SRO parameters on the actual spectra is beyond the typical resolution of Mössbauer spectroscopy.

ii)   The spectra are only sensitive to the average SRO parameter in the first two neighbor shells, defined by $\alpha = (8\alpha_1 + 6\alpha_2) / 14$ [2], and not separately to $\alpha_1$ and $\alpha_2$. This is because each satellite peak sums up contributions from configurations with the same total number of Cr neighbors $m = m_1 + m_2$. Thus, ordering effects are reflected on the spectra only as an average over the two neighbor shells.

Taking into account the above observations it is seen that the configuration probabilities $P_{m_1 m_2}$ can be effectively modeled in the PVM with only two parameters: the alloy concentration $c$ and the averaged SRO parameter $\alpha$. This represents a significant reduction in the number of free parameters compared to previous methods and it contributes to the robustness of the analysis. Furthermore, more information is provided as the alloy concentration is also obtained. On the other hand, we stress here that the sensitivity of Mössbauer spectroscopy is limited for detailed studies of SRO effects since only a single parameter related to ordering can in principle be obtained.

## V.    EXPERIMENTAL

Three high purity Fe-Cr model alloys with designations Fe-5Cr, Fe-10Cr, and Fe-15Cr, of Cr atomic concentration 0.058, 0.108 and 0.150, respectively, were obtained from the European Fusion Development Agreement (EFDA). The alloys were prepared in the form of cylindrical bars by induction melting and contain very low levels of impurities. Their final heat treatment was 1h at $750-850\,^{\circ}C$ under pure Ar flow followed by air cooling. Slices of 0.5 mm thickness were cut from the bars, cold rolled to a thickness of approx. 70 μm and chemically cleaned to remove surface contamination. A separate batch of specimens was annealed at 800°C for 8 hours under vacuum ($1\times10^{-6}$ mbar) and then rapidly withdrawn from the furnace.  Finally all Mössbauer samples were brought to a final thickness between 40 and 50 μm by chemical polishing using a solution containing oxalic acid (2.5 wt. %), adipic acid (4.7 wt. %), $NH_4HF_2$(4.4 wt. %) and $H_2O_2$(36 vol.  %). The Cr concentration in the prepared specimens was checked by energy dispersive X-ray analysis (EDX) under a scanning electron microscope (SEM). The EDX-measured concentration was in agreement with the nominal one within the resolution of the EDX method.

Mössbauer spectroscopy was carried out using a spectrometer in constant acceleration mode with a $^{57}Co(Rh)$ source. The data were collected in transmission mode at 77 K. The spectra were analyzed by standard non-linear least-squares methods based on eq. (10) with configuration probabilities calculated by means of the PVM. Except from the alloy concentration $c$ and the average SRO parameter $\alpha$, other parameters that are refined during fitting include: the HMF shifts $\Delta H_0$, $\Delta H_1$ and $\Delta H_2$; the Gaussian width $\sigma$ of the lineshape [cfg. eq. (8)] which is taken equal for all terms; the relative intensity of the 2nd and 5th line in the six-line pattern which may vary depending on the direction of magnetization. The lorentzian width $\gamma$ of the nuclear resonances is estimated from calibration measurements utilizing a pure iron foil. Source broadening and spectrometer resolution effects are effectively incorporated in the value of $\gamma$. The isomer shift $\delta$, measured with respect to pure iron, was considered as linearly correlated with the HMF, $\delta_{m_1m_2}=\delta_0+\delta_1H_{m_1m_2}$, as suggested in [34], and the parameters $\delta_0$ and $\delta_1$ were co-refined during the fitting procedure.

## VI.   RESULTS AND DISCUSSION

Experimental Mössbauer absorption spectra of the annealed (AN) Fe-Cr alloys are shown in fig. 3 as a function of source velocity. The spectra could be successfully fitted by the PVM model and the resulting curves are also shown in fig. 3 together with the fitting residuals. The model reproduces the experimental behavior very well; however, there are some systematic discrepancies mainly at the peak positions. This may be due to details of the Mössbauer lineshape, e.g., due to finite absorber thickness, that the Voigt profile does not correctly account for. The results of the analysis for the Cr concentration, the average SRO parameter and the relative HMF shifts are listed in Table II. Error values reported in the table correspond to the statistical uncertainty resulting from the least-square minimization. The experimental spectra of the specimens in the cold-worked (CW) state exhibit marginal differences from the AN ones. They were analyzed in exactly the same manner and the corresponding results are also listed in Table II.

We first discuss the hyperfine parameters deduced from the analysis. In order to compare our results obtained at 77 K with previous experiments performed typically at room temperature (RT) or with ab-initio calculations which in principle correspond to 0 K, the HMF shifts are given in Table II normalized to the HMF of the host. The magnitude of $H_{Fe}$ at liquid-nitrogen temperature is 339 kG [36], approximately 2.5% higher than at RT. Values from previous experimental work quoted in Table II have been scaled to the appropriate value of $H_{Fe}$ at the corresponding measurement temperature. Theoretical calculations cited in Table II give directly the relative shifts $\Delta H / H_{Fe}$.

As seen from Table II the obtained HMF shifts per Cr nearest and next nearest neighbor, $\Delta H_1$ and $\Delta H_2$, are equal in the CW and AN state of each sample, within experimental error. Thus the metallurgical condition does not affect these parameters. The negative sign of $\Delta H_1$ and $\Delta H_2$ reflects the fact that Cr atoms, coupling anti-ferromagnetically to the host lattice, tend to reduce the magnetic moment on neighboring iron atoms. Compared to previous measurements in the Fe-Cr system [12, 13], the observed relative shifts are found slightly larger in magnitude, by about 1% on the average. This is probably due to the lower temperature of measurement. Earlier Mössbauer studies of HMF shifts as a function of temperature for a number of transition metal impurities in iron found variations of the same order of magnitude [37]. Notably, the value of $\Delta H_1 / H_{Fe} = -10.2\%$ obtained for the lowest concentration of Cr is in excellent agreement with the first-principles calculations of Dederichs at al.

[23] and Rahman et al. [24], which both correspond to the low-temperature and low-concentration limit. The value of $\Delta H_2 / H_{Fe}$ is slightly below the theoretical predictions. With increasing Cr concentration $\Delta H_1 / H_{Fe}$ is found to increase in magnitude. This is not in accordance with previous work of Dubiel & Cieslak [13] who found a constant $\Delta H_1$ in Fe-Cr alloys with Cr concentration up to 25 at. %.

The shift in the HMF of iron atoms with no Cr neighbors, $\Delta H_0$, is positive and between 2 and 4 % of the host HMF depending on concentration. The positive sign of $\Delta H_0$ has been attributed to an increase of the magnetic moment of isolated iron atoms which partly compensates the large decrease of magnetic moment around Cr solutes. $\Delta H_0$ is the only parameter that is affected by the metallurgical processes as it is found systematically larger in the annealed samples. This behavior has been previously observed by Dubiel & Zukrowsky [12]. Our $\Delta H_0$ data are in good agreement with their results, both qualitatively and quantitatively. The increase of $\Delta H_0$ during annealing is possibly related to the relaxation of stresses that are present in the CW specimens.

In all samples the coefficient $\delta_1$ that couples the HMF to isomer shifts is found equal to $(5.7 \pm 0.3) \times 10^{-4}$ mm s$^{-1}$ kG$^{-1}$. The corresponding changes in isomer shift per Cr atom may be calculated from the data of Table II as $\delta_1 \times \Delta H_1 = -0.021 \pm 0.001$ mm s$^{-1}$ and $\delta_1 \times \Delta H_2 = -0.013 \pm 0.001$ mm s$^{-1}$ for nearest and next nearest Cr neighbors, respectively. These values are in good agreement with previous work [13].

Summarizing the discussion on hyperfine parameters, our findings are in good agreement with previous Mössbauer investigations of Fe-Cr alloys with the sole exception for the concentration dependence of $\Delta H_1$. This shows that the use of the PVM for determining the probabilities of different Cr configurations does not interfere with the estimation of the hyperfine parameters related to these configurations. It remains an open question if the observed concentration dependence of $\Delta H_1$ is due to the PVM, i.e., if it is a real effect brought forward by the new method of analysis or if it is just an artifact introduced by this method.

We now turn to discuss the parameters relevant to the alloy configuration, i.e., the concentration and the SRO parameter, extracted on the basis of the PVM. In the Fe-5Cr alloy the concentration predicted by the Mössbauer analysis in both CW and AN specimens is in very good agreement with the nominal one, within statistical error. A negative SRO parameter is found indicating an ordering tendency of the solute atoms. For both metallurgical conditions the same value of $\alpha$ is obtained, within errors. This is in contrast to the results of ref. [13] where $\alpha$ is reported to even change sign when the sample is subjected to metallurgical processes very similar to the ones employed here. The discrepancy is probably due to the limitations of the method employed in [13] for the estimation of $\alpha$ (cfg. section IV and [14]). In Fe-10Cr the Mössbauer-estimated concentration in both CW and AN samples is found slightly lower than the nominal one; however, differences are comparable to the statistical error. The SRO parameter is found negative in Fe-10Cr as well, both in the CW and AN state. Finally, in Fe-15Cr the concentration obtained from the fitting procedure is 12 at.%, i.e., 3 at.% lower than the nominal one, which is a substantial deviation. This occurs in both CW and AN samples. The SRO of Fe-15Cr is found again negative in both metallurgical conditions and of about the same magnitude as in the other alloys.

Thus, the picture that emerges from the Mössbauer analysis is that (i) the apparent concentration is in agreement with the nominal one at low concentration but is found gradually lower, starting at about 10 at.% Cr; (ii) that the SRO parameter is negative and of the same magnitude throughout the studied concentration range. The first of these observations is puzzling since the Mössbauer specimens were also tested with EDX and it was confirmed that their average Cr concentration is in accordance with the nominal one. Further, the observed SRO behavior is different from previous DNS results [1, 2]. Both techniques obtain the same average SRO parameter of $\alpha \cong -0.03$ for low Cr concentration. However, the sign inversion found by DNS as the concentration is increased is not confirmed by the present results. Thus, it appears that discrepancies arise with the Mössbauer-estimated alloy properties at high Cr concentration. In the following we discuss possible causes for these discrepancies, which may either lie within our method of analysis or they may base on real physical phenomena.

First, the convergence behavior of our least-squares procedure was extensively tested, in order to ensure that it was not trapped at some shallow minimum within the parameter space. The tests showed that in all cases the least-square solutions represent true minima and that there is sufficient sensitivity in determining the values of concentration and SRO parameter. This is exemplified in fig. 4 for Fe-15Cr. In fig. 4(a) the Cr concentration is changed by $\pm 0.5$ at.% around the fitted value of 12 at.% and the corresponding spectra are plotted together with the actual least-squares solution and the experimental data. Only the left-most peak of the Moessbauer pattern is shown for more detail. As seen in the figure, the deviation of the simulated curves from the experimental spectrum is clearly visible already at this small level of parameter change. The same is true for the SRO parameter which is varied by $\pm 0.01$ around the fitted value of $-0.03$. The corresponding simulated spectra are shown in fig. 4(b) and the sensitivity of the fitting procedure concerning the value of $\alpha$ is confirmed.

We now discuss the assumptions of our model, which have been put forward in section III. The restriction to only Cr atoms in the first two coordination shells has been previously employed by many investigators [9, 12, 13] and, moreover, it is supported by first-principles theoretical calculations [23, 24]. Theoretical justification exists also for our second assumption, i.e., that configuration probabilities are mainly determined by pair correlations. Effective interactions in Fe-Cr of atomic clusters other than the pair, e.g., of three-atom or four-atom clusters, have been generally found much weaker than pair interactions in calculations of configurational energy by several authors [3, 26, 30]. Consequently, they will result in small corrections to the PVM-calculated $P_{m_1 m_2}$. The only assumption that does not have theoretical justification yet is the additivity of HMF shifts and one may expect that some refinements are necessary in this respect. It is noted that atomic configurations involving more than one Cr atom, where additivity is really probed, become important in the same concentration region where the discrepancies of Mössbauer results occur. Thus, more work is needed, both theoretical and experimental, to clarify this point.

On the other hand, a possible interpretation of our findings could be offered by Cr clustering or segregation in the concentrated alloys. The Fe-Cr phase diagram [38] exhibits a miscibility gap separating the Fe-rich alpha phase from the Cr-rich alpha-prime phase. This leads to alloy decomposition by the formation of alpha-prime precipitates in super-cooled alloys with Cr concentration above about 10 at. % aged at temperatures between ~400 and 550ºC [38]. Thus, it could be argued that the concentrated alloys of the present study are in an early stage of decomposition where not well-defined precipitates but small Cr-enriched regions may have formed. These regions may exhibit a range of concentrations within the miscibility gap. Support for this assumption is given by preliminary small-angle neutron scattering measurements on these alloys in the as-received state, which have indicated the existence of a fine dispersion of nanometer-sized objects [39]. With regard to the Mössbauer effect, the high Cr concentration regions will have significantly reduced HMF values according to eq. (9). The presence of a large number of closely spaced satellites will result in an almost structureless Mössbauer pattern; moreover, the overall intensity of the pattern will be reduced due to the low iron content in these regions. A model calculation based on equations (9)-(10), assuming that the 15 at. % Cr alloy has decomposed into high concentration regions of 50 at. % Cr and low concentration regions of 12 at. % Cr , shows that the high concentration regions indeed contribute an almost flat component under the main six-line pattern, with an intensity of about 5 % relative to the central peaks. Due to its low intensity and lack of structure this component may be lost in the non-resonant transmission background of the experimental Mössbauer spectrum. This could actually explain the discrepancies observed here between the Mössbauer-estimated and nominal alloy concentrations. What actually happens is that Mössbauer spectroscopy probes the concentration within the Cr-depleted regions of the specimens which is apparently lower than the nominal. The value of 12 at.% Cr obtained by the Mössbauer analysis in Fe-15Cr compares well with the solubility limit of Cr in the alpha phase.

For essentially the same reasons discussed in the previous paragraph, the Mössbauer-estimated SRO parameter is related only to the Cr-depleted alpha phase regions. This may also account for the discrepancy regarding the SRO behavior between our results and those obtained by DNS. The latter technique measures an average over the entire sample volume occupied by both Cr-depleted and Cr-rich phases. At the low concentration of 5.8 at.% Cr, where the alloy is single phase alpha, both methods yield exactly the same $\alpha$. At higher concentrations, where Cr clustering effects become significant, DNS measures the average SRO parameter which increases towards positive values, while Mössbauer spectroscopy sees only the SRO of the Cr-depleted alpha phase which remains almost constant. Similar arguments have been also put forth by Erhart et al. [4] in their recent simulations of ordering and precipitation in Fe-Cr. They interpreted the SRO inversion observed by DNS as being the result of an averaging of the SRO parameters in the alpha and alpha-prime phases. Notably, they find that in the alpha phase the SRO parameter is always negative and does not vary appreciably with concentration, which is in accordance with our findings.

## VII. CONCLUSIONS

A novel approach is presented for analyzing Mössbauer spectra of short-range ordered Fe-Cr alloys. It relies on the Probability Variation Method (PVM) of Clapp [15, 16] for obtaining the probabilities of different Cr configurations around a Mössbauer-active iron atom. Within the PVM, the *a priori* assumption is made that these configuration probabilities are fully determined by pair correlations or, equivalently, by the Warren-Cowley short range order (SRO) parameters. The physical background of this assumption is that the configurational energy of the alloy is determined mainly by effective pair interactions. This approximation is quite justified for Fe-Cr as shown by recent theoretical investigations. Regarding the hyperfine parameters in Fe-Cr, typical assumptions are adopted as employed in many previous studies, i.e., that only Cr atoms in the first two shells create measurable perturbations to hyperfine parameters and that these perturbations are additive when multiple Cr neighbors are present.

The advantage of the new method is that information on SRO is obtained directly through the least-squares fitting of experimental spectra, providing also an estimate of the alloy concentration. Further, the number of adjustable parameters is thereby significantly reduced making the analysis much more robust.

The new method of analysis has been first tested on synthesized spectra obtained from Monte-Carlo simulated Fe-Cr configurations with known degrees of short-range order. The tests have shown (i) that the PVM estimates correctly the relevant Cr configuration probabilities and (ii) that by application of the new method on the simulated spectra the SRO parameter and the concentration are accurately obtained. With the tests it was also possible to asses the sensitivity of Mössbauer spectroscopy in determining the SRO parameters. It is established that the method is sensitive to ordering effects; however, the information that can be extracted from the spectra is limited to a single parameter, namely, the average SRO parameter over the first two nearest neighbor coordination shells.

Finally, experimental spectra from three Fe-Cr alloys of well characterized purity and composition (5.8, 10.8 and 15.0 at. % Cr) could be successfully analyzed with the new method. The obtained hyperfine parameters are in good agreement with previous Mössbauer investigations of Fe-Cr alloys. The Cr concentration estimated from the Mössbauer spectra is in agreement with the nominal one for concentrations up to about 10 at. % Cr. However, for the alloy with the highest nominal concentration of 15.0 at. % Cr the Mössbauer analysis estimates only 12.0 at. %. Further, the Mössbauer-estimated SRO parameter is found negative in the whole concentration range. The SRO parameter obtained for low Cr concentration is in agreement with diffuse neutron scattering (DNS) results, but at higher concentration the sign inversion observed previously by DNS is not confirmed. These discrepancies occurring at higher Cr concentrations could be attributed to alloy decomposition which is known to occur in Fe-Cr at these concentration levels. However, it cannot be excluded that the discrepancies are due to the assumption of the additivity of HMF shifts caused by Cr neighbors, which has been typically employed so far in the analysis of Mössbauer spectra and is also adopted in this work. The additivity would be expected to start failing as more Cr atoms are first and second neighbors to Fe, which occurs as the Cr concentration in the alloy increases.

## ACKNOWLEDGMENTS

Dr. E. Devlin is greatly acknowledged for performing the Mössbauer measurements and for fruitful discussions. This work has been partly supported by the European Fusion Development Agreement (EFDA) through its Contract of Association with the Hellenic Republic.

## REFERENCES

[1] I. Mirebeau, M. Hennion, and G. Parette. First measurement of short-range-order inversion as a function of concentration in a transition alloy. *Phys. Rev. Lett.*, 53 (7): 687–690, Aug 1984. doi: 10.1103/PhysRevLett.53.687.

[2] I. Mirebeau and G. Parette. Neutron study of the short range order inversion in $Fe_{1-x}Cr_x$ . *Phys. Rev. B*, 82 (10): 104203, Sep 2010. doi: 10.1103/PhysRevB.82.104203.

[3] M. Yu. Lavrentiev, R. Drautz, D. Nguyen-Manh, T. P. C. Klaver, and S. L. Dudarev. Monte carlo study of thermodynamic properties and clustering in the bcc fe-cr system. *Phys. Rev. B*, 75 (1): 014208, Jan 2007. doi: 10.1103/PhysRevB.75.014208.

[4] Paul Erhart, Alfredo Caro, Magdalena Serrano de Caro, and Babak Sadigh. Short-range order and precipitation in fe-rich fe-cr alloys: Atomistic off-lattice monte carlo simulations. *Phys. Rev. B*, 77 (13): 134206, Apr 2008. doi: 10.1103/PhysRevB.77.134206.

[5] P. Olsson, IA Abrikosov, L. Vitos, and J. Wallenius. Ab initio formation energies of fe–cr alloys. *J. Nucl. Mater.*, 321 (1): 84–90, 2003. doi: 10.1016/S0022-3115(03)00207-1.

[6] P. Olsson, IA Abrikosov, and J. Wallenius. Electronic origin of the anomalous stability of fe-rich bcc fe-cr alloys. *Phys. Rev. B*, 73 (10): 104416, 2006. doi: 10.1103/PhysRevB.73.104416.

[7] A Alamo, M Horsten, X Averty, E.I Materna-Morris, M Rieth, and J.C Brachet. Mechanical behavior of reduced-activation and conventional martensitic steels after neutron irradiation in the range 250–450°c. *J. Nucl. Mater.*, 283–287, Part 1 (0): 353 – 357, 2000. ISSN 0022-3115. doi: 10.1016/S0022-3115(00)00076-3.

[8] G. K. Wertheim, V. Jaccarino, J. H. Wernick, and D. N. E. Buchanan. Range of the exchange interaction in iron alloys. *Phys. Rev. Lett.*, 12 (1): 24–27, Jan 1964. doi: 10.1103/PhysRevLett.12.24.

[9] I. Vincze and IA Campbell. Mossbauer measurements in iron based alloys with transition metals. *J. Phys. F: Met. Phys.*, 3: 647, 1973. doi: 10.1088/0305-4608/3/3/023.

[10] I. Vincze and AT Aldred. Mössbauer measurements in iron-base alloys with nontransition elements. *Phys. Rev. B*, 9 (9): 3845, 1974.

[11] SM Dubiel and K. Krop. Influence of neighbouring chromium atoms on hyperfine fields at $^{57}$Fe nuclei and isomer shifts in Fe-Cr alloys. *Le Journal de Physique Colloques*, 35 (C6): 6–6, 1974. doi: 10.1051/jphyscol:1974695.

[12] S. M. Dubiel and J. Zukrowski. Mössbauer effect study of charge and spin transfer in fe-cr. *J. Magn. Magn. Mater.*, 23 (2): 214 – 228, 1981. ISSN 0304-8853. doi: 10.1016/0304-8853(81)90137-2.

[13] S. M. Dubiel and J. Cieslak. Short-range order in iron-rich Fe-Cr alloys as revealed by Mössbauer spectroscopy. *Phys. Rev. B*, 83 (18): 180202, May 2011. doi: 10.1103/PhysRevB.83.180202.

[14] LH Schwartz and A. Asano. Determination of local atomic order using the Mössbauer effect. *J. Phys. Colloques*, 35 (C6): 6–6, 1974. doi: 10.1051/jphyscol:1974694.

[15] Philip C. Clapp. Theoretical determination of n-site configuration probabilities from pair correlations in binary lattices. *J. Phys. Chem. Solids*, 30 (11): 2589 – 2598, 1969. ISSN 0022-3697. doi: 10.1016/0022-3697(69)90267-4.

[16] Philip C. Clapp. Atomic configurations in binary alloys. *Phys. Rev. B*, 4 (2): 255–270, Jul 1971. doi: 10.1103/PhysRevB.4.255.

[17] E. Fermi. Über die magnetischen Momente der Atomkerne. *Z. Phys. A*, 60 (5): 320–333, 1930. doi: 10.1007/BF01339933.

[18] Brent Fultz. *Characterization of Materials*, chapter Mössbauer Spectrometry. John Willey, 2011.

[19] M. B. Stearns. Internal magnetic fields, isomer shifts, and relative abundances of the various fe sites in fesi alloys. *Phys. Rev.*, 129: 1136–1144, Feb 1963. doi: 10.1103/PhysRev.129.1136.

[20] L. H. Schwartz and D. Chandra. Hyperfine fields in concentrated Fe-Cr alloys. *Phys. Status Solidi B*, 45 (1): 201–208, 1971. ISSN 1521-3951. doi: 10.1002/pssb.2220450122.

[21] Mary Beth Stearns. Spin-Density Oscillations in Ferromagnetic Alloys. I. "Localized" Solute Atoms: Al, Si, Mn, V, and Cr in Fe. *Phys. Rev.*, 147: 439–453, Jul 1966. doi: 10.1103/PhysRev.147.439.

[22] Mary Beth Stearns and Stephen S. Wilson. Measurements of the conduction-electron spin-density oscillations in ferromagentic alloys. *Phys. Rev. Lett.*, 13: 313–315, Aug 1964. doi: 10.1103/PhysRevLett.13.313.

[23] PH Dederichs, B. Drittler, R. Zeller, H. Ebert, and W. Weinert. Calculation of hyperfine interaction parameters in metals-friedel oscillations in dilute fe and cu alloys. *Hyperfine Interact.*, 60 (1): 547–562, 1990. doi: 10.1007/BF02399824.

[24] Gul Rahman, In Gee Kim, H. K. D. H. Bhadeshia, and Arthur J. Freeman. First-principles investigation of magnetism and electronic structures of substitutional 3$d$ transition-metal impurities in bcc Fe. *Phys. Rev. B*, 81: 184423, May 2010. doi: 10.1103/PhysRevB.81.184423.

[25] D.D. Fontaine. Cluster approach to order-disorder transformations in alloys. *Solid state physics*, 47: 33–176, 1994. doi: 10.1016/S0081-1947(08)60639-6.

[26] A V Ruban and I A Abrikosov. Configurational thermodynamics of alloys from first principles: effective cluster interactions. *Reports on Progress in Physics*, 71 (4): 046501, 2008. doi: 10.1088/0034-4885/71/4/046501.

[27] Bernd Schönfeld. Local atomic arrangements in binary alloys. *Progress in Materials Science*, 44 (5): 435 – 543, 1999. ISSN 0079-6425. doi: 10.1016/S0079-6425(99)00005-5.

[28] DMC Nicholson, RI Barabash, GE Ice, CJ Sparks, J. Lee Robertson, and C. Wolverton. Relationship between pair and higher-order correlations in solid solutions and other ising systems. *J. Phys.: Condens. Matter*, 18: 11585, 2006. doi: 10.1088/0953-8984/18/50/013.

[29] A. V. Ruban, P. A. Korzhavyi, and B. Johansson. First-principles theory of magnetically driven anomalous ordering in bcc fe-cr alloys. *Phys. Rev. B*, 77: 094436, Mar 2008. doi: 10.1103/PhysRevB.77.094436.

[30] P. A. Korzhavyi, A. V. Ruban, J. Odqvist, J.-O. Nilsson, and B. Johansson. Electronic structure and effective chemical and magnetic exchange interactions in bcc fe-cr alloys. *Phys. Rev. B*, 79: 054202, Feb 2009. doi: 10.1103/PhysRevB.79.054202.

[31] Rafa Idczak, Robert Konieczny, and Jan Chojcan. Atomic short-range order in iron based Fe-Mo alloys studied by [57]Fe Mössbauer spectroscopy. *Hyperfine Interact.*, pages 1–6, 2011. ISSN 0304-3843. doi: 10.1007/s10751-011-0458-6. 10.1007/s10751-011-0458-6.

[32] V.V. Ovchinnikov, B.Yu. Goloborodsky, N.V. Gushchina, V.A. Semionkin, and E. Wieser. Enhanced atomic short-range ordering of the alloy fe-15 at. % cr caused by ion irradiation at elevated temperature and thermal effects only. *Appl. Phys. A*, 83 (1): 83–88, April 2006. doi: 10.1007/s00339-005-3458-z.

[33] P. C. Gehlen and J. B. Cohen. Computer simulation of the structure associated with local order in alloys. *Phys. Rev.*, 139 (3A): A844–A855, Aug 1965. doi: 10.1103/PhysRev.139.A844.

[34] D.G. Rancourt and J.Y. Ping. Voigt-based methods for arbitrary-shape static hyperfine parameter distributions in Mössbauer spectroscopy. *Nucl. Instr. Meth. B*, 58 (1): 85 – 97, 1991. ISSN 0168-583X. doi: 10.1016/0168-583X(91)95681-3.

[35] G Apostolopoulos. *PVMOS Manual - Probability Variation Method for Mössbauer spectroscopy*. NCSR "Demokritos", 2012. URL http://www.ipta.demokritos.gr/gapost/software.htm

[36] P. C. Riedi. Temperature Dependence of the Hyperfine Field and Hyperfine Coupling Constant of Iron. *Phys. Rev. B*, 8: 5243–5246, Dec 1973. doi: 10.1103/PhysRevB.8.5243.

[37] I. Vincze and G. Grüner. Temperature Dependence of the Hyperfine Field at Iron Atoms near 3$d$ Impurities. *Phys. Rev. Lett.*, 28: 178–181, Jan 1972. doi: 10.1103/PhysRevLett.28.178.

[38] VP Itkin. Phase diagrams of binary iron alloys. *ASM International, Materials Park, OH*, page 102, 1993.

[39] K Mergia, G Apostolopoulos, and S Messoloras. SANS investigation of model FeCr alloys. Annual Report 38, Association EURATOM-Hellenic Republic, 2009. URL http://www.hellasfusion.gr/-images/stories/Annexes/2009/Annex38.pdf.

**Table I.** Results of the analysis of Monte-Carlo (MC) simulated Mössbauer spectra of Fe-Cr alloys; $c_{Cr}$ denotes the Cr concentration, $\alpha$ is the average short-range order parameter in the first two nearest neighbor shells, and $\Delta H_1 / H_{Fe}$ and $\Delta H_2 / H_{Fe}$ denote the relative shifts in the hyperfine magnetic field of Fe per Cr nearest and next nearest neighbor, respectively.

| | $c_{Cr}$ (at. %) | $\alpha$ | $\Delta H_1 / H_{Fe}$ (%) | $\Delta H_2 / H_{Fe}$ (%) |
|---|---|---|---|---|
| random alloy [Fig. 1(b)] | | | | |
| MC input | 15 | 0 | -10 | -7 |
| least-squares fit | 14.83 | -0.004 | -9.99 | -7.03 |
| ordering alloy [Fig. 1(c)] | | | | |
| MC input | 15 | -0.05 | -10 | -7 |
| least-squares fit | 14.79 | -0.059 | -9.94 | -7.00 |
| clustering alloy [Fig. 1(d)] | | | | |
| MC input | 15 | 0.05 | -10 | -7 |
| least-squares fit | 15.10 | 0.056 | -10.01 | -6.98 |

**Table II.** Results of the analysis of experimental Mössbauer Spectra. $c_{Cr}$ denotes the Cr concentration, $\alpha$ is the average short-range order parameter in the first two nearest neighbor shells, and $\Delta H_1 / H_{Fe}$ and $\Delta H_2 / H_{Fe}$ denote the relative shifts in the hyperfine magnetic field of Fe per Cr nearest and next nearest neighbor, respectively.

| Alloy | Condition | Nominal $c_{Cr}$ (at. %) | Mössbauer Results | | | | |
|---|---|---|---|---|---|---|---|
| | | | $c_{Cr}$ (at. %) | $\alpha$ | $\Delta H_0 / H_{Fe}$ (%) | $\Delta H_1 / H_{Fe}$ (%) | $\Delta H_2 / H_{Fe}$ (%) |
| Fe-5Cr | CW* | 5.8 | 5.4(2) | -0.030(4) | 2.66(3) | -10.2(1) | -7.0(1) |
| | AN | | 5.6(2) | -0.032(5) | 3.02(1) | -10.2(1) | -6.9(1) |
| Fe-10Cr | CW | 10.8 | 9.8(3) | -0.028(6) | 3.83(4) | -10.8(1) | -6.7(1) |
| | AN | | 9.6(3) | -0.037(6) | 3.95(3) | -10.9(1) | -6.9(1) |
| Fe-15Cr | CW | 15.0 | 12.1(4) | -0.031(10) | 3.76(13) | -11.3(1) | -7.1(2) |
| | AN | | 12.0(4) | -0.030(10) | 3.88(4) | -11.3(1) | -6.9(1) |
| Average values | | | | | 3.6(4)[1] | -10.8(4) | -6.9(1) |
| Dubiel & Cieslak [14] | | | | | | -9.39(15) | -6.45(20) |
| Dubiel & Zukrowski [13] | | | | | 3.8(1.0)[1] | -9.56(40) | 6.4(4) |
| Dederichs et al. [22] | | | | | | -10.0 | -7.6 |
| Rahman et al. [23] | | | | | | -9.9 | -8.0 |

*CW: Cold-worked; AN: Annealed.

[1]Average over AN samples

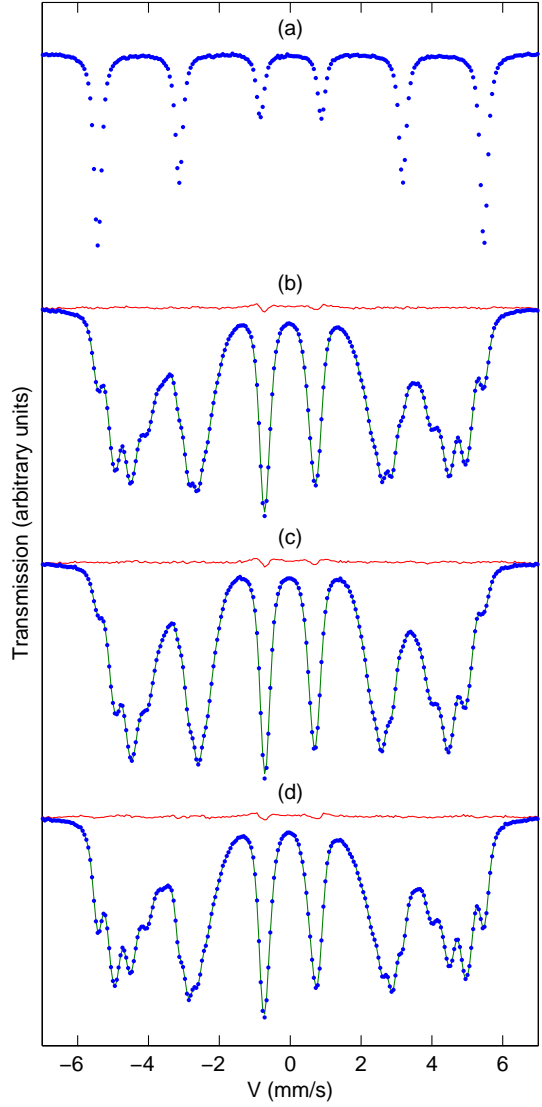

FIG. 1 (Color Online) Monte-Carlo simulated Mössbauer spectra (blue dots) for (a) pure Fe and a Fe-15 at. % Cr alloy with SRO parameter (b) $\alpha = 0$, (c) $\alpha = -0.05$, and (d) $\alpha = 0.05$. Continuous lines (green) are least-squares fits as described in section IV; the fitting residuals are also shown. The alloy parameters estimated by the fitting are listed in Table I.

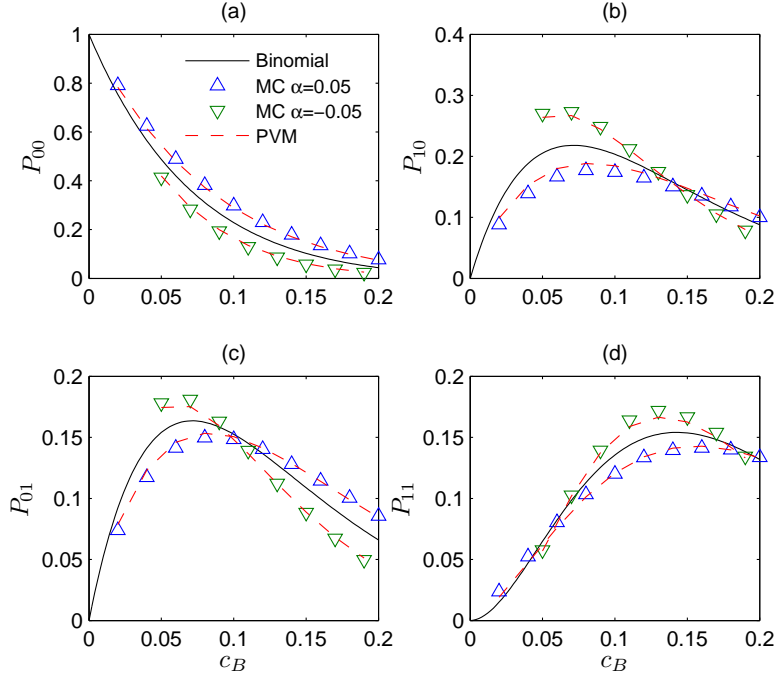

FIG 2 (Color Online) For a binary alloy A-B the probability $P_{m_1 m_2}$ is shown as a function of B concentration, $c_B$, for the occurrence of $m_1$ first-order and $m_2$ second-order nearest neighbors of type B around a central atom of type A. Black continuous curves represent the $P_{m_1 m_2}$ for a random alloy as given by the binomial distribution. Upper triangles (blue) and down triangles (green) are calculated by the Monte-Carlo (MC) technique for a short-range ordered alloy with Warren-Cowley SRO parameters $\alpha = \alpha_1 = \alpha_2 = 0.05$ and $\alpha = \alpha_1 = \alpha_2 = -0.05$, respectively. Higher order SRO parameters are set equal to 0. The red dashed curves correspond to $P_{m_1 m_2}$ obtained by the Probability Variation Method (PVM) of Clapp for the same Warren-Cowley SRO parameters.

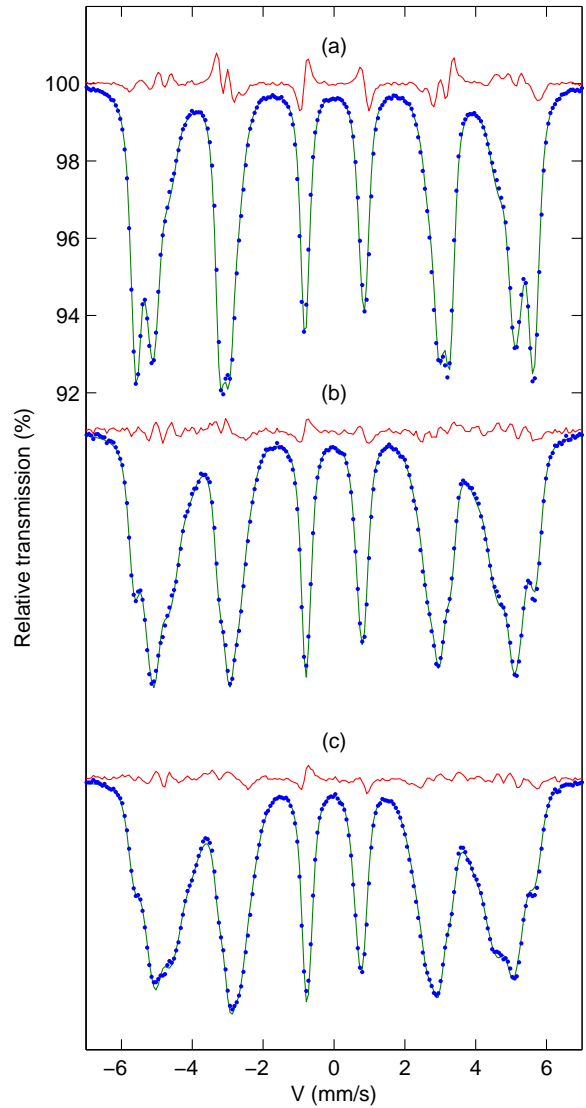

FIG 3. (Color Online) Moessbauer transmission spectra (blue dots) measured at 80 K as a function of source velocity for three Fe-Cr alloys with Cr concentration (a) 5.8 at.%, (b) 10.8 at.% and (c) 15.0 at. %. Continuous lines (green) represent fits to the experimental data as described in the text. The difference between measurement and fit is also shown (red).

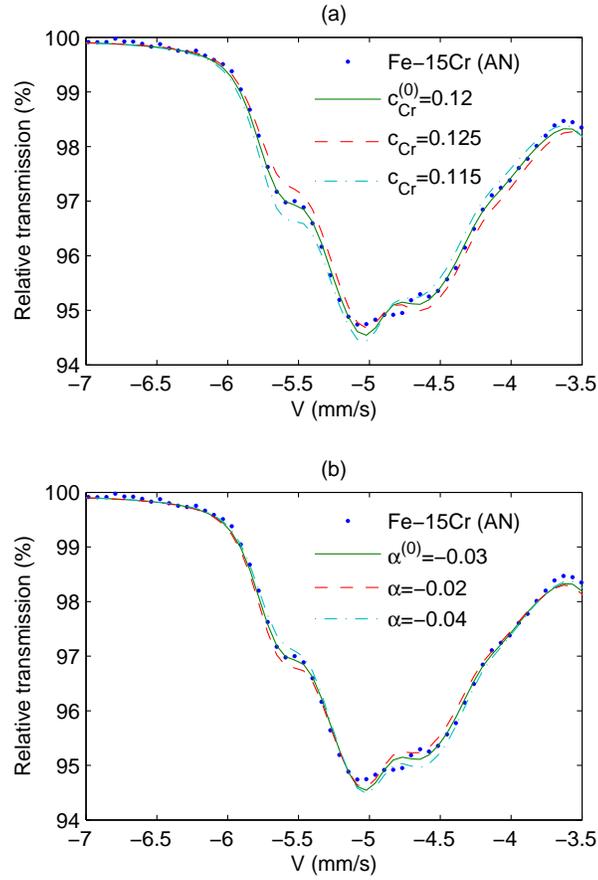

FIG. 4 (Color online) Sensitivity of the least-square solution on the fitting parameters. The dots and the green continuous curve on both (a) and (b) depict experimental data and least-square fit, respectively, for the left-most Moessbauer line of the annealed (AN) Fe-15Cr alloy. In (a) the value of the Cr concentration is changed from the fitted value of $c_{Cr}^{(0)} = 0.120$ to 0.125 (red dashed curve) and then to 0.115 (cyan dash-dot curve); all other spectrum parameters are kept constant. In (b).the value of the SRO parameter is changed from the fitted value of $\alpha^{(0)} = -0.03$ to $-0.02$ (red dashed curve) and then to $-0.04$ (cyan dash-dot curve).